\documentstyle[spie,epsf]{article} 
\input{psfig}   

\title{
Plans for a 10-m Submillimeter-wave Telescope\\
at the South Pole
}

\author{Antony A. Stark\supit{a}, John E. Carlstrom\supit{b}, Frank P. Israel\supit{c}, Karl M. Menten\supit{d},\\
Jeffrey B. Peterson\supit{e}, T. G. Phillips\supit{f}, Giorgio Sironi\supit{g} and Christopher K. Walker\supit{h}
\skiplinehalf 
\supit{a}Smithsonian Astrophysical Observatory, 60 Garden St. MS78, Cambridge,
MA 02138, USA
\skiplinehalf 
\supit{b}University of Chicago, Department of Astronomy and Astrophysics, \\
5640 S. Ellis Ave.,  Chicago, IL 60613, USA
\skiplinehalf 
\supit{c}Sterrewacht Leiden, Postbus 9513, 2300 RA, Leiden, the Netherlands
\skiplinehalf 
\supit{d}Max-Planck-Institut f\"{u}r Radioastronomie, Auf dem H\"{u}gel 69, 53121 Bonn, Germany
\skiplinehalf 
\supit{e}Carnegie Mellon University, Department of Physics, Pittsburgh, PA 15213-3890,
USA
\skiplinehalf 
\supit{f}Downs Laboratory of Physics, California Institute of Technology, Pasadena, CA 91125, USA
\skiplinehalf 
\supit{g} Universit\'{a} Degli Studi di Milano, Dipartimento di Fisica, \\
Sezione di Astrofisica Relativistica e Cosmologia, Via Celoria 16, 20133 Milano, Italy
\skiplinehalf 
\supit{h}Steward Observatory, University of Arizona, Tucson, AZ 85721, USA
}

\authorinfo{Other author information: \\
A.A.S. (correspondence): E-mail: {\tt aas@cfa.harvard.edu}; WWW: {\tt http://cfa-www.harvard.edu/$\sim$aas/tenmeter/tenmeter.html}}

\begin{document} 
  \maketitle 

\def\aj{{AJ}}			
\def\araa{{Ann. Rev. Astron.\& Astrophys.}}		
\def\apj{{ApJ}}			
\def\apjl{{ApJ (Letters)}}		
\def\apjs{\ref@jnl{ApJS}}		
\def\ao{\ref@jnl{Appl.Optics}}		
\def\apss{\ref@jnl{Ap\&SS}}		
\def\aap{\ref@jnl{A\&A}}		
\def\aapr{\ref@jnl{A\&A~Rev.}}		
\def\aaps{\ref@jnl{A\&AS}}		
\def\azh{\ref@jnl{AZh}}			
\def\baas{\ref@jnl{BAAS}}		
\def\jrasc{\ref@jnl{JRASC}}		
\def\memras{\ref@jnl{MmRAS}}		
\def\mnras{{MNRAS}}		
\def\pra{\ref@jnl{Phys.Rev.A}}		
\def\prb{\ref@jnl{Phys.Rev.B}}		
\def\prc{\ref@jnl{Phys.Rev.C}}		
\def\prd{\ref@jnl{Phys.Rev.D}}		
\def\prl{\ref@jnl{Phys.Rev.Lett}}	
\def\pasp{\ref@jnl{PASP}}		
\def\pasj{\ref@jnl{PASJ}}		
\def\qjras{\ref@jnl{QJRAS}}		
\def\skytel{\ref@jnl{S\&T}}		
\def\solphys{\ref@jnl{Solar~Phys.}}	
\def\sovast{\ref@jnl{Soviet~Ast.}}	
\def\ssr{\ref@jnl{Space~Sci.Rev.}}	
\def\zap{\ref@jnl{ZAp}}			
\let\astap=\aap
\let\apjlett=\apjl
\let\apjsupp=\apjs
\def\deg{\hbox{$^\circ$}}
\def\sun{\hbox{$\odot$}}
\def\earth{\hbox{$\oplus$}}
\def\la{\mathrel{\hbox{\rlap{\hbox{\lower4pt\hbox{$\sim$}}}\hbox{$<$}}}}
\def\ga{\mathrel{\hbox{\rlap{\hbox{\lower4pt\hbox{$\sim$}}}\hbox{$>$}}}}
\def\sq{\hbox{\rlap{$\sqcap$}$\sqcup$}}
\def\arcmin{\hbox{$^\prime$}}
\def\arcsec{\hbox{$^{\prime\prime}$}}
\def\fd{\hbox{$.\!\!^{\rm d}$}}
\def\fh{\hbox{$.\!\!^{\rm h}$}}
\def\fm{\hbox{$.\!\!^{\rm m}$}}
\def\fs{\hbox{$.\!\!^{\rm s}$}}
\def\fdg{\hbox{$.\!\!^\circ$}}
\def\farcm{\hbox{$.\mkern-4mu^\prime$}}
\def\farcs{\hbox{$.\!\!^{\prime\prime}$}}
\def\fp{\hbox{$.\!\!^{\scriptscriptstyle\rm p}$}}
\def\micron{\hbox{$\mu$m}}
\def\onehalf{\hbox{$\,^1\!/_2$}}	
\def\onethird{\hbox{$\,^1\!/_3$}}
\def\twothirds{\hbox{$\,^2\!/_3$}}
\def\onequarter{\hbox{$\,^1\!/_4$}}
\def\threequarters{\hbox{$\,^3\!/_4$}}
\def\ubvr{\hbox{$U\!BV\!R$}}		
\def\ub{\hbox{$U\!-\!B$}}		
\def\bv{\hbox{$B\!-\!V$}}		
\def\vr{\hbox{$V\!-\!R$}}		
\def\ur{\hbox{$U\!-\!R$}}		
% greater than twiddle and less than twiddle
\def\gs{\mathrel{\raise0.35ex\hbox{$\scriptstyle >$}\kern-0.6em 
\lower0.40ex\hbox{{$\scriptstyle \sim$}}}}
\def\ls{\mathrel{\raise0.35ex\hbox{$\scriptstyle <$}\kern-0.6em 
\lower0.40ex\hbox{{$\scriptstyle \sim$}}}}
\newcount\lecurrentfam
\def\LaTeX{\lecurrentfam=\the\fam \leavevmode L\raise.42ex
\hbox{$\fam\lecurrentfam\scriptstyle\kern-.3em A$}\kern-.15em\TeX}
\def\plotone#1{\centering \leavevmode
\epsfxsize=\textwidth \epsfbox{#1}}
\def\plottwo#1#2{\centering \leavevmode
\epsfxsize=.45\textwidth \epsfbox{#1} \hfil
\epsfxsize=.45\textwidth \epsfbox{#2}}
\def\plotfiddle#1#2#3#4#5#6#7{\centering \leavevmode
\vbox to#2{\rule{0pt}{#2}}
\includegraphics{#1}}
%%%%%%%%%%%%%%%%%%%%%%%%%%%%%%%%%%%%%%%%%%%%%%%%%%%%%%%%%%%%% 
\begin{abstract}
A 10 meter diameter submillimeter-wave telescope has been proposed for
installation and scientific use at the NSF Amundsen-Scott South Pole Station.  
Current evidence indicates that the South Pole 
is the best submillimeter-wave telescope site 
among all existing or proposed ground-based observatories.
Proposed scientific programs place stringent requirements on the 
optical quality of the telescope design.  In particular, reduction of the 
thermal background and offsets requires an off-axis, unblocked aperture, and the large 
field of view needed for survey observations requires shaped optics.  
This mix of design elements is well-suited for large-scale 
(square degree) mapping of line and continuum radiation from submillimeter-wave 
sources at moderate spatial resolutions (4 to 60 arcsecond beam size) 
and high sensitivity (milliJansky flux density levels).
The telescope will make
arcminute angular scale, high frequency Cosmic Microwave Background 
measurements from the best possible ground-based site, using an 
aperture which is larger than is currently possible on orbital or 
airborne platforms.
The telescope design is homologous.  Gravitational changes in pointing and 
focal length will be accommodated by active repositioning of the secondary 
mirror.  The secondary support, consisting of a large, enclosed beam, permits 
mounting of either a standard set of Gregorian optics, or prime focus 
instrumentation packages for CMBR studies.  A tertiary chopper is located at 
the exit pupil of the instrument.
An optical design with a hyperboloidal primary mirror and a concave secondary mirror
provides a flat focal surface.   The relatively large classical aberrations present in
such an optical arrangement can be small compared to diffraction at submillimeter
wavelengths.
Effective use of this telescope will require development of large (1000 element)
arrays of submillimeter detectors which are background-limited when 
illuminated by antenna temperatures near 50 K.
\end{abstract}

\keywords{Antarctic, South Pole, submillimeter, astronomy, telescopes}
%%%%%%%%%%%%%%%%%%%%%%%%%%%%%%%%%%%%%%%%%%%%%%%%%%%%%%%%%%%%%
  \section{INTRODUCTION} 

Of all ground-based observatory sites that have been 
tested for submillimeter-wave sky quality, the best results have
come from the Amundsen-Scott South Pole Station.
In the current decade, a year-round observatory has been established
at the Pole by the Center for Astrophysical Research in Antarctica (CARA),
an NSF Science and Technology Center.  
CARA operates three major telescope facilities:  AST/RO (the Antarctic
Submillimeter Telescope and Remote Observatory, a 1.7-m telescope\cite{stark97a}),
Python (a Cosmic Microwave Background experiment), and SPIREX (the South Pole
Infrared Explorer, a 60-cm telescope).
These facilities are conducting site characterization and astronomical 
investigations from millimeter wavelengths to the near-infrared\cite{lane98}.
Profiles of temperature, pressure, and water vapor above the Pole have been 
measured at least daily for decades by the South Pole 
meteorology office\cite{schwerdtfeger}.
Measured water vapor values are extremely low because the air is 
dessicated by frigid temperatures (annual average: -49C, median
winter PWV: 0.25mm).  Winds at the Pole are unusually low (maximum
recorded wind speed over 30 years: 25 m/s), 
and rain is completely absent\cite{schwerdtfeger}.
Atmospheric opacity at the Pole has been routinely measured at 225 GHz and 
492 GHz using skydip techniques\cite{lane98,chamberlin94,chamberlin95,chamberlin97}.  
The 225 GHz results are comparable to
those made at the proposed NRAO Millimeter Array site at Atacama, Chile.
This, together with the significantly higher atmospheric pressure at the
Pole, necessarily implies that the PWV and submillimeter-wave opacity at 
the Pole are somewhat better than at Atacama.

A wide-field, off-axis 10 m diameter submillimeter-wave telescope 
has been proposed for the South Pole
by an international consortium.  
%Participating institutions
%at this time are: Smithsonian Astrophysical Observatory, 
%the University of Chicago, the
%California Institute of Technology, Carnegie Mellon University, Leiden Observatory,
%Max-Planck-Institut f\"{u}r Radioastronomie, and Steward Observatory.
The South Pole 10 m (SP 10m) telescope will be a powerful and highly flexible 
instrument, usable for a wide variety of submillimeter observations in 
observational cosmology,
star formation, chemistry and dynamics of the interstellar medium, 
galactic structure, and solar system studies. 
The SP 10m telescope will be optically superior to existing
telescopes with respect to 
scattered radiation, cross-polarization, and field of view and will be 
located at a site which is consistently much more transparent in 
the submillimeter-wave atmospheric windows than any other existing or
proposed observatory site\cite{lane98}.  It will therefore be orders 
of magnitude faster than existing instruments and capable
of detecting much weaker sources. It 
will greatly enhance submillimeter-wave capability in the Southern Hemisphere 
and provide a single-dish complement to the Smithsonian Astrophysical
Observatory (SAO) Submillimeter Array (SMA) and the 
NRAO Millimeter Array (MMA).
It will make use of large, sensitive bolometer and heterodyne array detectors
which are advanced versions of those currently in use on the Caltech Submillimeter
Observatory (CSO) and the James Clerk Maxwell Telescope (JCMT).
The 10 meter telescope will be able to feed 1000 pixels in a detector array.

%Proposed scientific programs place particularly stringent requirements on the 
%optical quality of the telescope design.  In particular, reduction of the 
%thermal background requires an off-axis, unblocked aperture, and the large 
%field of view needed for survey observations requires shaped optics.  
%This mix of design elements is particularly well suited for large-scale 
%(square degree) mapping of line and continuum radiation from submillimeter-wave 
%sources at moderate spatial resolutions (4 to 60 arcsecond beam size) 
%and high sensitivity (milliJansky flux density levels). 

\section{Science Goals for the SP 10m}

A workshop was held at Harvard University on 28 March 1997 to discuss
science goals for the SP 10m project.  
Some critical projects that drive the telescope design are:  

\paragraph{Primary cosmic microwave background anisotropy at arcminute scales.}
At angular scales near 0.5\deg , standard cosmological models predict 
relatively large anisotropy caused by an acoustic ``bounce" that occurred before
recombination.
The spatial frequency at which this peak occurs
depends on the fundamental cosmological parameter
$\Omega$.
Additional smaller peaks are expected at multiples of the
peak spatial frequency. 
Present CMBR observations show a clear maximum in the anisotropy spectrum
at a spatial scale somewhere near 1\deg , but the spectrum lacks detail
and the harmonic peaks at smaller scales have not yet been seen.
The SP 10m will study the damping tail region at arcminute scales,
providing a strong test
of the acoustic oscillation model and providing measurements of peak positions
that will augment the lower spatial frequency information obtained with 
meter-class telescopes.  Data from the SP 10m will be particularly vital if
the universe turns
out to be open, and the main spatial frequency peak occurs at scales near 0.1\deg .
To study primary anisotropy at arcminute angular scales, it will be important
to simultaneously understand the secondary contribution to anisotropy due to
Sunyaev-Zel'dovich (S-Z) distortion in galaxy  clusters between the
recombination last scattering
surface and the observer.  Because of the low sky noise available at the Pole,
the SP 10m can be used over a wide range of frequencies, including the
200-300 GHz range of frequencies required to allow spectral separation of
thermal S-Z effect from primary anisotropy.  Since there are 
$10 ^ 4$ to $10 ^ 5$ visible clusters of galaxies, both the small beam size 
and wide spectral range of the SP 10m
will be needed to locate the cluster-free regions of sky that
will be observed for primary anisotropy studies.

\paragraph{Sunyaev-Zel'dovich effect.}
As CMB photons travel from the surface of last scattering
to the observer, secondary anisotropies can arise due to the
interaction of the CMB photons with intervening matter.
Of particular interest is the S-Z effect, which
occurs when CMB photons travel through a cluster of galaxies\cite{Rephaeli95}.
Approximately $10\,$\% of the total mass of rich clusters of galaxies
is in the form of hot ($\sim 10^8\,$K) ionized plasma.
Compton scattering of CMB photons by electrons in this intra-cluster
plasma can result in an optical depth as high as 0.02, resulting in
a distortion of the CMB spectrum at the mK level.

There are two components of the S-Z effect which result from the two distinct
velocity components of the scattering electrons. The thermal component
is due to
the thermal (random) velocities of the scattering electrons.  The kinematic
component is due to the bulk velocity of the intra-cluster 
gas with respect to the rest
frame of the CMB.  
%In Figure~\ref{fig:szspectra}, the spectral distortion
%produced by each  of the two components is shown.  
%As evident from the
%figure, 
The
two S-Z components have distinct spectra which can be separated by
observations at millimeter wavelengths.

The thermal component of the S-Z effect can be used in
combination with X-ray data to provide a measure of the Hubble
constant ($H_0$). In addition, when combined with a measurement of
electron temperature, the ratio of the kinematic
and thermal component amplitudes provides a direct measurement of the cluster's
peculiar velocity relative to the rest frame of the CMB.
The observed surface brightness difference of both the thermal and kinematic
components is independent of the cluster redshift, as long as the cluster is
resolved.  
Using the SP 10m, accurate S-Z measurements can be made
throughout the Universe, all the way back to the epoch of formation of
the hot intra-cluster gas.
This will be an important new probe of universal structure formation processes: 
a sample of peculiar velocities throughout the Hubble volume.

\paragraph{Continuum detection of high redshift protogalaxies not detectable in the 
visible or near-IR.} The majority of 
the luminosity resulting from the collapse energy of galaxies and 
the first generations of stars may not appear at visual or 
near-infrared wavelengths (even in the rest frame of the distant 
galaxy), but may instead be reradiated by dust\cite{meurer98,pearson96}.
Submillimeter-wave point sources with no bright visible counterpart
may already have been detected\cite{smail98}.
In the broadband rest-frame spectrum of normal galaxies are two 
roughly equal flux peaks: visible light produced by stars, and 
far-infrared radiation near 100\micron \ produced by dust 
and the fine-structure cooling lines of the interstellar medium. 
To understand the nature of protogalaxies, it will be necessary to 
observe them at submillimeter, near-infrared, and visible 
wavelengths.
The energetics of protogalaxy formation will not be understood
until there is a deep, high-resolution survey of the submillimeter-wave
background\cite{longair93,blain96a,blain96b,hughes97a,hughes97b}.

\paragraph{Line detection of high redshift protogalaxies not detectable in the 
visible or near-IR.}
The ${}^2 P_{3/2} \rightarrow {}^2 P_{1/2}$
line of [${\rm C \, \scriptstyle II \, }$] at $\lambda = 158 \micron$
is the brightest emission line in the spectrum of most galaxies.
From the radio to X-ray, the wavelength of highest flux density in a 
galaxian spectrum is usually the  peak of this line; 
as much as 0.5\% of the total luminosity of a galaxy can be emitted in
the single spectral line\cite{stacey91,wright91}.  
The [${\rm C \, \scriptstyle II \, }$] line-to-continuum ratio
is high in low metallicity systems, so that line searches for protogalaxies may be
more efficient than continuum searches \cite{petrosian69,loeb93,stark97c}.
Measurement of the [${\rm C \, \scriptstyle II \, }$] line yields
chemical and kinematic information for galaxies detected by other
means.

\paragraph{Continuum survey of cloud cores.}
An unbiased wide-area continuum survey at $\lambda = 350\micron $ 
of all nearby molecular regions to identify
dense cloud cores should be able to locate all such regions within 
500 pc of the Sun.  
This project can best be carried out at a 
wavelength of  $\simeq$ 350\micron;
observing at this relatively long wavelength
will ensure that even the most heavily obscured (deeply
embedded) sources will be included.  
The SP 10m will be 
particularly capable of tracing
extended protostellar disks, and connections between these disks and
surrounding, placental material.
Follow-on studies at 200 \micron \  wavelength may help
discriminate among protostellar models.

\paragraph{A spectral line survey of detected cloud cores
in kinematic tracers of the star-forming gas.}  
Such a survey would enable discrimination among the variety of complex motions found
in these regions, and in particular, identification of objects in which infall
of material is occurring.

\begin{table}[tbh!]
\begin{center}
\caption[Continuum Sensitivity of Submillimeter Telescopes] 
{{\bf Continuum Sensitivity of Submillimeter Telescopes\ \ } 
\\
\label{table:continuum} }
\begin{tabular}{lrrrrrrrrr}
Telescope & {$A^{\rm a}$} & {$R^{\rm b}$}& 
            {$S^{\rm c}$}& \multicolumn{3}{c}{NEFD$^{\rm d}$} 
            & \multicolumn{3}{c}{Time in hours to survey} \\
& $({\rm m^2})$ & $( '')$ &  & \multicolumn{3}{c}{$({\rm mJy \, s^{1/2}})$} & 
             \multicolumn{3}{c}{1 square degree at 1 mJy} \\
& & &  & $850 \mu \rm m$ & $450 \mu \rm m$ & $350 \mu \rm m$ & $850 \mu \rm m$ & $450 \mu \rm m$ & $350 \mu \rm m$ \\
%\tableline
\\
SP 10 m & 79 & 11 & 38 & {\it 60} & {\it 64} & {\it 74} & {\it 95} & {\it 108} & {\it 144}\\
SP 30 m & 711 & 4  & 4 & {\it 7} & {\it 7} & {\it 8} & {\it 11} & {\it 12} & {\it 16}\\
AST/RO & 2 & 65 & 92 & {\it 2160} & {\it 2300} & {\it 2660} &  ${\it 5.1 \times 10^4}$ & ${\it 5.8 \times 10^4}$ & ${\it 7.7 \times 10^4}$\\
JCMT & 177 & 7 & 5 & 80 & 700 & {\it 760} & $1.3 \times 10^3$ & $9.8 \times 10^4$& ${\it 1.2 \times 10^5}$\\
CSO & 79 & 11 & 11 & {\it 150} & 2000 & {\it 2200} & {$\it 2.0 \times 10^3$} & $3.6 \times 10^5$ & ${\it 4.4 \times 10^5}$\\
SOFIA & 5 & 44 & 50 &  & {\it 200} & {\it 200} &  & {\it 800} & {\it 800}\\
FIRST & 7 & 32 & 4.3 &  & {\it 54} & {\it 54} &  & {\it 678} & {\it 583}\\
SMA & 227 & 2 & 0.14 & {\it 134} & {\it 1142} & {\it 1250} & ${\it 3.6 \times 10^4}$ & ${\it 9.3 \times 10^6}$ & ${\it 1.9 \times 10^7}$ \\
MK array$^{\rm e}$ & 483 & 0.5 & 0.02 & {\it 59} & {\it 719} & {\it 790} & ${\it 4.9 \times 10^4}$ & ${\it 2.6 \times 10^7}$ & ${\it 5.2 \times 10^7}$ \\
``small'' MMA & 2010 & 0.2 & 0.08 & {\it 7} & {\it 45} & {\it 52} & ${\it 230}$ & ${\it 3.4 \times 10^4}$ & ${\it 7.5 \times 10^4}$\\
``large'' MMA & 7000 & 0.2 & 0.06 & {\it 2} & {\it 12} & {\it 15} & ${\it 19}$ & ${\it 2.4 \times 10^3}$ & ${\it 6.4 \times 10^3}$

\end{tabular}
\end{center}
\noindent 
{
Notes: 
\hfill\break
{a.}\,{Telescope area ($\rm m^2$)} \hfill
\break
{b.}\,{Resolution element ($\rm arcsec$) for $\lambda = 450 \micron$.
Resolution element scales as $\lambda$.} \hfill
\break
{c.}\,{Instantaneous sky coverage ($\rm arcmin^2$) for $\lambda = 450 \micron$.
Instantaneous Sky Coverage 
scales as $\lambda^2$ for interferometers,
is independent of $\lambda$ for most single-dish instruments, and is 4.3,
5.0 and 2.7 arcmin$^2$ at $480\micron$, $350\micron$ and $250\micron$,
respectively,
for FIRST (G. Pilbratt, personal communication).  }
\hfill
\break
{d.}\,{Noise Equivalent Flux Density, the sensitivity to point sources whose
positions are known.
Numbers in italics are predicted sensitivities; numbers not in italics are
measured, on-the-telescope values and are subject to downward revision with
improved techniques.  Predicted sensitivities are optimistic in the
sense that in all cases they are near the thermal background limit, a
limit that has not yet been achieved in practical
submillimeter-wave bolometer systems.
This table is based on the work of Hughes and Dunlop\cite{hughes97a}.}
\hfill
\break
{e.}\,{Mauna Kea array consisting of the SMA, the CSO and the JCMT}
}
\end{table}

\paragraph{Thermal balance of star-forming regions:  
cooling line inventory and PDR studies.}
Theoretical studies have established the importance of molecular
cooling in the initial phases of star formation; 
cooling at densities $\ge$ $10^6 {\rm cm}^{-3}$  is by a large number of lines
of many different species.  

\paragraph{Observations of the ionized nitrogen line near 1.46 THz (205 \micron).} 
The winter South Pole sky becomes 30\% transparent at this frequency.
The [${\rm N \, \scriptstyle II \, }$] line 
will be a powerful probe in
conjunction with observation of 
[${\rm C \, \scriptstyle II \, }$] from SOFIA and
[${\rm C \, \scriptstyle I \, }$].
These observations can be used to help determine the astrophysically 
important carbon to oxygen ratio and are relevant for studies of the Milky Way
and external galaxies. 

\paragraph{Polarization studies.} 
The optical design of the 10 m telescope makes it 
particularly well-suited for measurement of extended polarized 
submillimeter-wave radiation. During the past decade, it has been 
demonstrated that measurement of linear polarization of the thermal 
emission from interstellar dust grains yields unique information 
concerning the configuration of magnetic fields in star-forming 
regions\cite{hildebrand95}.  Theoretical work suggests 
that these fields may determine the masses of the stars that 
form\cite{shu87}. 
In regions with strong, 
ordered fields, the theory predicts that low-mass stars will 
form. When the fields are weaker, they are overwhelmed by gravity and 
higher mass stars result. 
The angular resolution and sensitivity of 
the South Pole 10 m telescope will allow polarimetric studies to be 
extended from high-mass star-forming regions (where magnetic 
fields have been studied in detail with airborne telescopes at 
far-infrared wavelengths and the large submillimeter telescopes on 
Mauna Kea), to the fainter, more extended regions where low-mass stars are born. 
The comparison will provide a crucial test for theories of star formation.

\paragraph{High resolution submillimeter-line studies of star-forming regions
in the Magellanic Clouds.}  The SP 10m provides an unparalleled opportunity to study star
formation in a low-metallicity system---perhaps an analog of star formation
in early galaxies.  These observations would be made in conjunction with the
MMA.

%\paragraph{Monitor the submillimeter flux of southern quasars and
%active galaxies.}  Many of these sources have large and variable submillimeter-wave
%fluxes.

\section{Telescope Sensitivity}

A 10 meter diameter telescope equipped with modern submillimeter-wave 
detectors and operating at the South Pole in winter is capable of 
detecting both line and continuum radiation from protogalaxies
at $z \sim 3$ in less than an hour.  
A protogalaxy which will evolve into a large (${\rm L^{*}}$) galaxy  
produces a continuum 
flux density of about 5 mJy and a [${\rm C \, \scriptstyle II} \, $] line power of about 
1 K ${\rm km~s^{-1}}$ at the focus of a 10 meter telescope\cite{stark97c}.
AST/RO has already detected submillimeter spectral lines having 0.5 K ${\rm km~s^{-1}}$ 
power\cite{stark97b}.  Even though 
this power level corresponds to a significantly larger flux in the beam of 
a 1.7 meter telescope, detection of these lines with AST/RO demonstrates
that an offset 10 meter antenna operating under the winter South Pole sky 
can overcome systematic noise problems at this power level and is capable of 
making observations of the requisite sensitivity for observing 
protogalaxies.  

Table \ref{table:continuum} shows the continuum
sensitivity and beam size of the SP 10m and other submillimeter-wave telescopes and
illustrates the strength of the SP 10m for deep continuum surveys.
The NEFD values in this table result from different assumptions for the different
instruments, assumptions which have varying degrees of certainly and 
practical verification; it should be understood that this table is speculative.
The high sensitivity per pixel for the SP 10m 
results from the combination of a relatively large
aperture, large available bandwidth, and fairly good sky; the fast survey
speed results from high sensitivity and a large field of view; 
a relatively small beam size provides freedom from confusion.
This combination of attributes assures that the SP 10m telescope will
make unique and important contributions to astronomy.

\begin{figure}[tb!] 
\begin{center} 
\leavevmode
\epsfxsize=7.0in
\epsffile{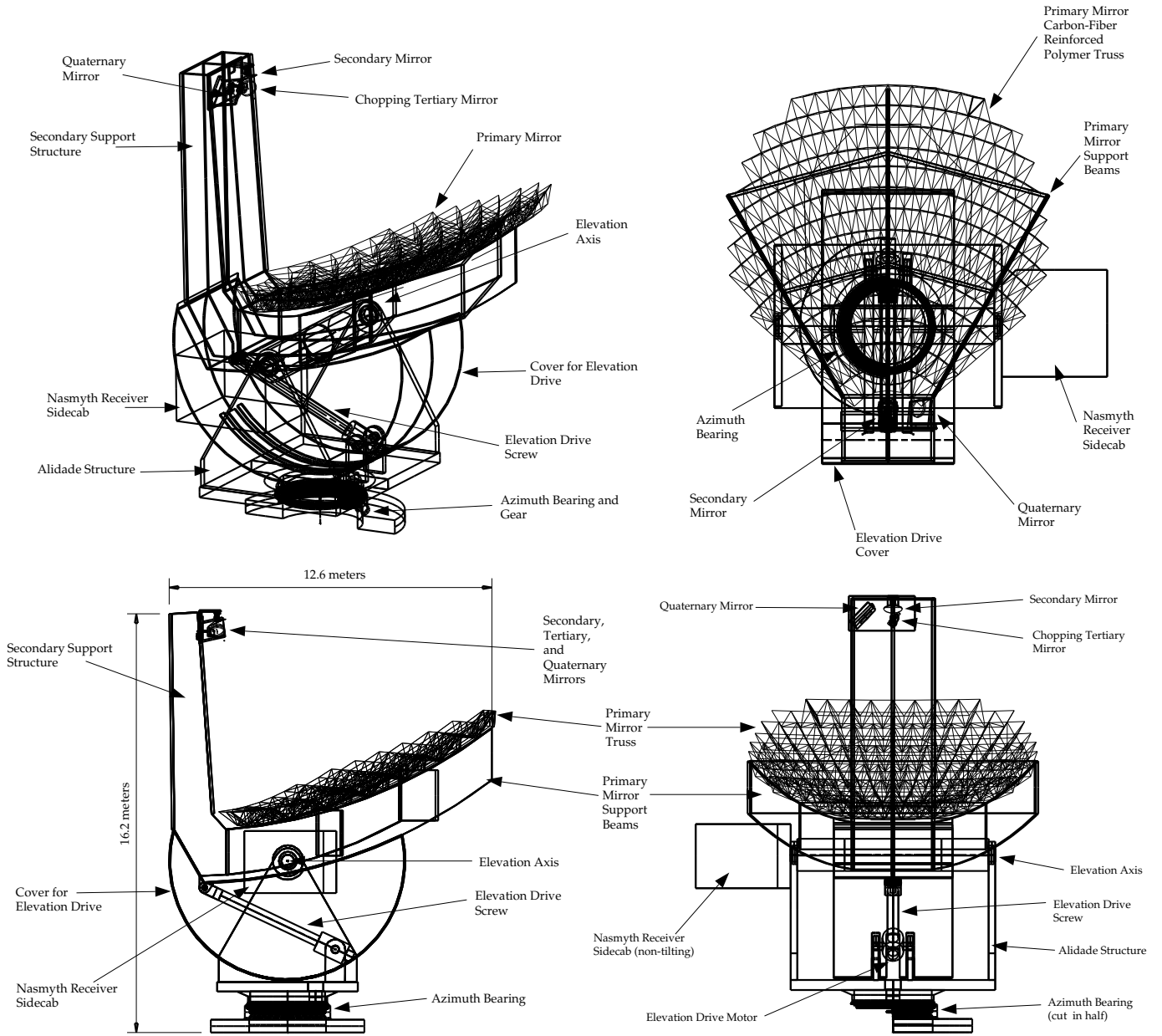}  \\
\end{center}
\caption[A Computer Model of the South Pole 10 meter Telescope.]
{{\bf A Computer Model of the South Pole 10 meter Telescope.\ \ }
\label{fig:sp10m} }
\end{figure}

\section{Design Concept}

A design study was conducted in 1996 by the SAO Central Engineering 
Department 
%of the Smithsonian Astrophysical Observatory (SAO) 
using funding contributed for that purpose by the Smithsonian Institution.  
That study has resulted in a preliminary 
design for the telescope, 
a static structural analysis of the design,
a detailed seven year budget, and
a plan for the construction of the telescope using the technology,
personnel, and facilities currently being used to construct the six
6-m diameter antennas of the 
%Smithsonian Astrophysical Observatory
Submillimeter Array (SMA).

%The planned observations 
%demand offset optics of at least 10 meters diameter, a surface accuracy of 
%12\micron \ RMS, and a field of view at least 30 resolution elements
%across.
%An instrument of this size and accuracy can be constructed 
%using machined aluminum panels, passively mounted on a 
%carbon-fiber-reinforced-polymer truss.  
%Shaped optical design will provide significant increase in field of view.  

%The effect of a submillimeter
%telescope or array on the South Pole infrastructure is roughly proportional to
%the number of antennas deployed and is independent of
%the size of those antennas.  Once set up and aligned at the South Pole, 
%the 10 meter telescope will not require power or other resources 
%significantly greater than AST/RO  
%does now; the power requirement for each 
%telescope is about 12 kW, most of it for cryogenic compressors.  
%A hypothetical South Pole antenna array, 
%however, would require 12 kW per antenna.  Optimal use of South Pole
%facilities argues for placing as much collecting area as possible in
%a single antenna with an aperture efficiency which is as high as
%possible, and with the greatest possible number of detectors looking
%at that aperture---these are the design rules followed in the
%SP 10m design.

Avoidance of internal reflections is critical to the design of 
submillimeter-wave telescopes.  Any receiver or detector placed at the 
focus of the instrument will necessarily emit into the telescope some 
amount of submillimeter-wave power in the band of interest.  If there is a 
reflection in the system, for example at the secondary mirror some 6 
meters distant,
then a resonant cavity is formed whose modes are spaced at 
roughly 25 MHz intervals,
and spurious features at those intervals will appear in the spectrometer whenever
there is a change in power level or cavity dimensions.  Submillimeter-wave
spectrometers cover more than 1 GHz of bandwidth, and a 25 MHz spurious
feature can masquerade as a typical astronomical or atmospheric line.
Also critical to submillimeter-wave telescope design is minimization of
variations in antenna thermal emissions as a function of chopper angle.
Chopper offsets can more easily be minimized in an off-axies design.

Dragone\cite{dragone82} has shown that if the offset angles in an offset 
Gregorian antenna are chosen correctly, then aberrations and 
cross-polarization effects in that offset 
antenna are the same as those in a conventional on-axis antenna with 
the same diameter
and focal length.  The beam efficiency, aperture efficiency, and 
sidelobe levels in the off-axis antenna
are better than those in the on-axis antenna, because in the on-axis design there
will be diffraction, reflection, and blockage from the secondary mirror 
and its supports.
An off-axis Gregorian telescope with correctly chosen offset angles
will always be optically superior to a similar on-axis configuration. 
An offset telescope allows for large prime-focus instrument packages of the type
used successfully on Python and other CMBR telescopes; these instruments {\em cannot}
be used with an on-axis design. 

\begin{table}[tbh!]
%\scriptsize
\caption[10m Telescope Design Parameters.]
{{\bf 10m Telescope Design Parameters.\ \ }
\label{table:parameters} }
\begin{tabular}{l c l}
\hline
optical performance       & & diffraction limited at 200 \micron \ wavelength\\
field of view             & & at least $20'$ diameter \\
tracking                  & & $1''$ maximum error in winds of $10{\rm ~m~s^{-1}}$ or less\\
slew speed                & & 0.5 rpm in winds of $30{\rm ~m~s^{-1}}$ or less\\
primary mirror homology   & & $< 12$ \micron \ RMS deviation from defined surface \\
primary aperture          & $d$ & 10000 mm\\
center of primary         & $\rm I_2$& (5300mm, 0, 1003.214mm)\\
paraxial focus of primary & $\rm F_1$ & (0, 0, 7000mm)\\
primary mirror materials  & & cast aluminum panels supported by \\
                          & & carbon-fiber-reinforced-polymer truss and titanium nodes\\
panel size                & & each approximately 0.75 square meters \\
length along central ray from primary\\ 
$~~~~$to secondary              & & 8403.375 mm\\
center of secondary       & $\rm I_1$& $(-265\rm mm , 0,7299.839\rm mm )$\\
first focus of secondary  & & (0, 0, 7000 mm)\\
second focus of secondary & $\rm F_0$& $(-11.16738 \rm mm ,0,-699.348 \rm mm)$\\
diameter of secondary     & &  700 mm\\
magnification of secondary& $m$ & $-20$\\
scaleless backfocal distance& $\beta$ & $0.1$\\
secondary materials       & & cast aluminum single panel\\
secondary mirror mount    & & actively-controlled hexapod\\
center of tertiary        & & $(-251.616\rm mm, 0, 6878.052\rm mm)$\\
chopping tertiary         & & opposed-torque balanced chopper \\
elevation drive system    & & counterweighted axis with ball screw drive or bull gear \\
azimuth drive system      & & opposed torque pinions on bull gear\\
azimuth bearing           & & fully-constrained gothic-arch with integral gear\\
encoder precision         & & 23 bits\\
drive system electronics  & & VME/VXI bus system \\
detector focal points     & & prime focus (for CMBR), Gregorian, Nasmyth\\
Nasmyth cabin size        & & 4 m $\times$ 4 m $\times$ 3 m\\
guide telescope           & & CCD camera at focus of 75 mm diameter f/9 lens
\end{tabular}
\end{table}

In the proposed telescope, all optical elements are
offset. The primary mirror
consists of passively-mounted aluminum panels on a carbon-fiber reinforced
polymer truss.  
%The truss is homologous, in the sense that gravitational 
%deflections in the primary mirror and primary mirror mount cause 
%distortion into a series of parabolic shapes with different pointings and
%focal lengths: these changes will be accommodated by active repositioning
%of the secondary mirror.  
The SP 10m will not use active surface technology
where the panels of the primary are repositioned automatically; the
primary is a passive structure which is manually adjusted.
The secondary support, consisting of a large 
enclosed beam, permits the mounting of both a concave secondary mirror
directed toward the Gregorian or Nasmyth focii
and a prime focus instrumentation package for CMBR studies.
A tertiary chopper is located at the exit pupil of the instrument, in a
design similar to that used successfully on the AST/RO telescope\cite{stark97a}.
The tertiary chopper can be repositioned to direct the beam to prime-focus
packages, or removed
to minimize the number of reflecting surfaces to the Gregorian focus.
All mechanical systems are enclosed to protect them from the
elements.  Waste heat from compressors and motors will be ducted inside this
enclosure.  Both axes are fully balanced with counterweights.

\paragraph{Homology of the Primary Mirror---}
An important aspect of this antenna design is use of homology, that is, the 
tendency for gravity-induced primary mirror surface deflections to act
as a focus shift.  Simply stated, gravitational deflections must be parabolic and 
predictable, permitting their effect to be removed by a pre-calculated 
repositioning of the secondary mirror.  Homologous primary mirror
design is a particularly effective technique for a South Pole telescope,
because gravitation rather than wind or temperature change is the dominant cause 
of structural deflections in the weather conditions prevalent at 
Pole\cite{schwerdtfeger,chamberlin97}.  A 
structural analysis has been made of the preliminary primary mirror design which 
includes the panels, 
back-up structure, steel support, secondary, and mounting arrangement. 
For the preliminary analysis, 
a single-sized structural component was 
used throughout the reflector support truss.  The analysis shows that 
the deflection in the primary resulting from a pointing change from zenith 
to an elevation angle of $50\deg $ follows a parabola to within 26 \micron ,
or 16 \micron \ RMS, with an absolute maximum deflection of 860 \micron .  This
is surprisingly good, considering that the preliminary design has equal-stiffness
truss elements and has not yet been optimized in any way.  
It therefore appears likely
that the 10 meter primary can be designed to be 
diffraction-limited at 200 \micron \ wavelength.
%for elevations between $90\deg$ 
%and $50\deg$.  
The next step in the design is to iteratively modify the 
stiffness of the carbon-fiber truss rods, in order to improve the homology.

In further design studies we will consider various 
primary mirror support arrangements, examining the effects of the placement 
of the elevation axis and the ball screw elevation drive. 
We will perform a detailed analysis of the truss structure elements, 
optimizing the primary structural response, further reducing the primary 
mirror deflections, while improving the homology.  A dynamical analysis of 
the primary mirror will consider the excitation of deflection modes by 
wind and by the drive system.

\section{Optical Design}

The optical design criteria  are:
\begin{enumerate}
\item{The field of view is diffraction-limited for
wavelengths as short as $\lambda = 200 \micron$.}
\item{The illumination of the primary does not change significantly
with chopper angle.}
\item{Large (1000 beam) focal plane arrays can be used in
a protected environment.}
\item{The field of view is large enough to 
accommodate these large focal-plane arrays while beam chopping.}
\end{enumerate}
These requirements differ from those of
visual-wavelength telescopes, where minimization of
blockage and coma typically lead to a Ritchey-Chretien design,
and also from those of most
radio telescopes, where maximization of  beam and aperture efficiency
for the central beam typically leads to a classical Cassegrain
configuration.  The SP 10m is novel because a large field of view
is needed in a situation where diffraction dominates the classical
aberrations; in this case maximizing the flatness of the focal
plane while providing a real exit pupil for the chopper location
leads to a modified offset Gregorian design. 
The optics are offset in order to eliminate blockage, and
the primary is shaped to enhance the field flatness.

\begin{figure}[tbh!]
\begin{center} 
\leavevmode
\epsfxsize=5.5in
\epsfbox{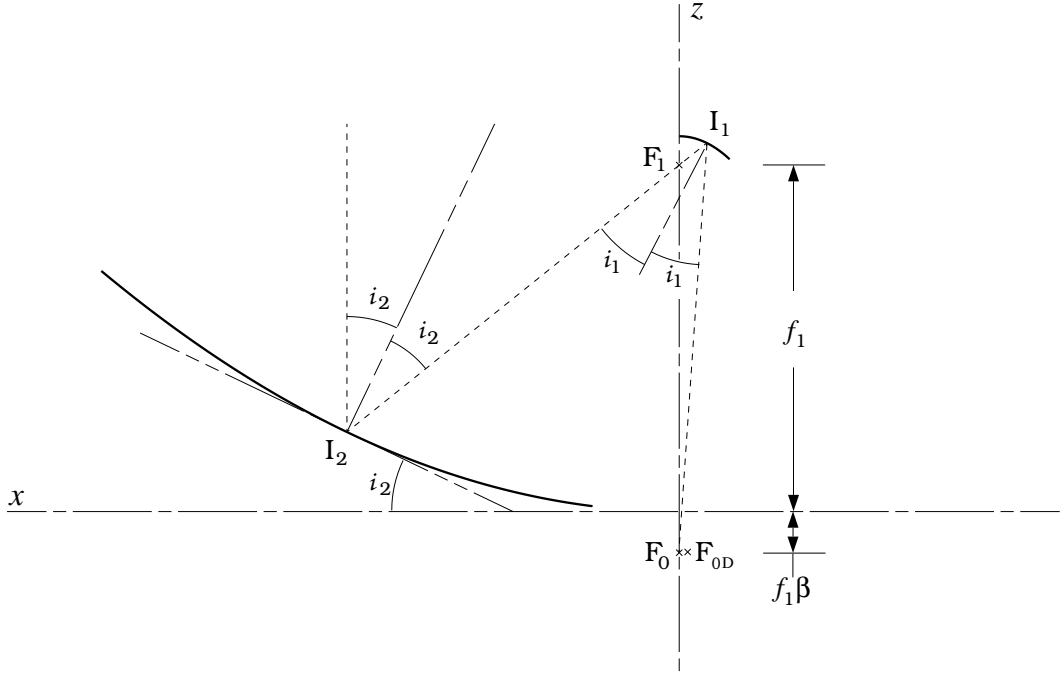}  \\
\end{center}
\caption[Scale drawing of the primary and secondary mirrors.]
{{\bf Scale drawing of the primary and secondary mirrors.\ \ }
The instrument focus indicated by ${\rm F_{0D}}$ satisfies the
Dragone\cite{dragone82} relation, Equation~\ref{eq:dragone} with 
$i_0 = 0$.  
The alternative focus
indicated by ${\rm F_{0}}$ nearly satisfies the Dragone relation
while greatly simplifying the calculation of aberrations, since
the line ${\rm F_1 F_0}$ is then the axis of symmetry of the mirrors,
and the mirrors are figures of revolution: the telescope is
then equivalent to an on-axis instrument with an off-axis aperture stop.
\label{fig:optics} }
\end{figure}

The primary
and secondary mirrors are symmetric about a vertical plane, and
Figure~\ref{fig:optics} shows the intersection of
this plane with the mirrors.
This optical design has two more degrees of freedom than an
on-axis telescope: the offset angles of the primary ($i_2$) and
secondary ($i_1$) mirrors.
Dragone\cite{dragone82} has shown that for an offset two mirror
system the aberration function is equivalent
to that of a single-mirror prime-focus paraboloid with offset angle $i_0$, where
\begin{equation}
\label{eq:dragone}
{\rm tan} ~ i_0 = (1-M){\rm tan} ~ i_1 + M~{\rm tan} ~ i_2, 
\end{equation}
where $M \equiv {\rm { (F_0 I_1)} \over (F_1 I_1)}$ is the off-axis 
magnification of the secondary.
%For the AST/RO Coud\'e focus, %$M = -30.03$, and $f_0 = 44,310 ~\rm mm$.  
Aberrations are minimized if ${\rm tan} ~ i_0 = 0$.  Substituting
this value into Equation~\ref{eq:dragone} couples the two offset
angles and reduces the
number of degrees of freedom by one.
An unfortunate effect of this relation is that 
the optical elements are then
no longer rotationally symmetric about the $z$ axis, greatly complicating
the analysis of system properties.  The final design of the optical system
will be accomplished through numerical techniques, so this is not a
fatal problem, but a simple analytic understanding of the effect of the
design constraints and parameters is needed in the early stages of the design. 
If $|M|$ is large, a simplifying assumption can be made.
The value of $i_1$ can be chosen 
so that the instrument focus lies on the $z$ axis (i.e., at ${\rm F_0}$ rather 
than ${\rm F_{0D}}$ in Figure~\ref{fig:optics}). 
The system is then equivalent to an on-axis optical
telescope with an off-axis aperture stop, and since that stop is large and
admits at least some rays from all radial zones, the aberrations of the
off-axis system are essentially the same as in the on-axis sytem. 
The value of $i_0$ 
%= - 0^\circ\!\!.7$, 
is still sufficiently small that the aberrations of the
resulting rotationally symmetric system are not very much
greater than for the fully optimized system. 
An analytic solution can therefore be obtained for a system
whose aberrations are only somewhat worse than those of the fully optimized design.
%Choosing $i_1$ this way is equivalent to Dragone's 
%minimum aberration condition ${\rm tan} ~ i_0 =0$ in the limiting case of large 
%secondary magnification $|M|$.  

Consider a Gregorian design, and parameterize the
mirror surfaces to allow deviations from the classical design
in order to optimize the optical performance.
An important practical constraint on this parameterization is that the primary be
a section of a figure of revolution, so that the mirror panels in every row are
identical.
The discussion below follows Schroeder\cite{schroeder87} in sign conventions 
and variable names.
In the classical case,
the secondary is a section of a prolate spheroid with
focii at $\rm F_0$ and $\rm F_1$, and the  primary is paraboloidal with a 
focus at $\rm F_1$ in Figure~\ref{fig:optics}.
The central ray is shown
as a short dashed line.  In this figure
the telescope is looking toward the zenith, along the $z$ axis.  
The coordinate system, $(x, y, z)$
has its origin a the vertex of the primary mirror paraboloid,
and the focus of the paraboloid is at $(0, 0, 7000 ~\rm mm)$, so the
surface of the primary mirror satisfies the equation
\begin{equation}
\label{eq:primary}
 0 =  x^2 + y^2  -  2 R_1 z + (1+K_1) z^2  , 
\end{equation}
where $R_1  = 2 f_1 =14000 ~\rm mm$ is the radius of curvature
at the vertex $(0, 0, 0)$, $f_1$ is the on-axis focal length,
and $K_1$ is the conic constant of the primary.
$K_1 = -1$ for a parabolic primary; this is the parameter which will be varied.
The primary mirror is a piece of this surface bounded by a cylinder
${\rm 10000 mm}$ in diameter, with its axis parallel 
to $z$ at $x = 5300~{\rm mm}, y = 0$.
The central ray follows this axis from $z = \infty $ until
it intercepts the center of the primary mirror 
at $(5300 ~\rm mm , 0, 1003.2 ~\rm mm)$,
where it reflects at an angle of incidence $i_2 \approx 25^\circ$,
through ${\rm F_1}$ at $(0, 0, 7000~\rm mm )$, to the center of the secondary.
The equation of the secondary can be written:
\begin{equation}
\label{eq:secondary}
 0 =  x^2 + y^2  -  2 R_2 (z - z_2)+ (1+K_2) (z - z_2)^2  , 
\end{equation}
where $R_2 $ is the on-axis radius of curvature (a negative
number for a concave secondary), $K_2$ is the conic constant
of the secondary, and $z_2$ is the distance between the
vertex of the secondary and the vertex of the primary.

The condition that the paraxial rays have focal points at
${\rm F_1}$ and ${\rm F_0}$ gives
\begin{equation}
\label{eq:R2}
R_2 = {{ m (1+\beta)}\over{m^2 - 1}} R_1
\end{equation}
and
\begin{equation}
\label{eq:z2}
z_2 = {{m - \beta}\over{2(m + 1)}} R_1,
\end{equation}
where $m$ is the magnification of the secondary and $\beta$
is the scaleless back focal length of the instrument, as
shown in Figure~\ref{fig:optics}.  The effective focal length of
the system is $f = m f_1$.  Since the system is
off-axis, the paraxial rays are not admitted by the aperture,
but rays hitting the inner part of the
mirror close to the $z$ axis are nearly paraxial.
Note that in a Gregorian system, $m$ is negative.  For
definiteness, choose $m = -20$ and $\beta = 0.1$, 
parameters which result in a reasonably small
secondary mirror and a useful distance between the secondary and
the focus ${\rm F_0}$.
Since $m$, $\beta$,
and $f_1$ are fixed, $R_1$, $R_2$ and $z_2$ are also fixed,
while the conic constants of the mirrors, $K_1$ and $K_2$, are varied.  

The condition that the focal point ${\rm F_0}$ be free of
spherical aberration gives a relation between the two
conic constants:
\begin{equation}
\label{eq:conic}
 K_2 = -\left({{m+1}\over{m-1}}\right)^2 +
{{m^3(m+1)}\over{(1+\beta)(m-1)^3}}(K_1 + 1) \equiv A + B (K_1 + 1).
\end{equation}
In the case of a classical Gregorian, the additional
requirement that the spherical aberration at ${\rm F_1}$ be zero
leads to a paraboloidal primary
with $K_1 = -1$, so $K_2 = A = -0.819$ and
the secondary is a prolate spheroid with focii at
${\rm F_1}$ and ${\rm F_0}$.  
For the present design, there is no advantage to a
sharp focus at ${\rm F_1}$, and no such requirement exists.
The system therefore has one
remaining degree of freedom, $K_1$, which can be chosen to 
minimize aberrations and maximize
image quality at  ${\rm F_0}$.

\paragraph{Aberrations---}
The classical third-order optical aberrations for rays reflecting within the telescope
can be written in terms of the parameters defined above\cite{schroeder87}.
Let $x$ be the $x$-coordinate value of the point where a ray strikes 
the primary mirror, and $\theta$ be the angle on the sky between that ray and the central
ray.  The worst-case ray strikes the outer edge of the primary, 
where $x = 1030 \rm mm$.
The angular tangential coma, $\theta_c$, is 
\begin{equation}
\label{eq:coma}
\theta_c = {{3 \theta x^2}\over{ 4 f^2}}\left[1 + {{m^2(m-\beta)}\over{2(1+\beta)}}
(K_1 + 1) \right] \equiv \theta {{x^2}\over{f^2}}[C + D (K_1 +1)].
\end{equation}
The angular astigmatism, $\theta_a$, is
\begin{equation}
\label{eq:astigmatism}
\theta_a = - {{ \theta^2 x}\over{f}}\left[{{m^2+\beta}\over{m(1+\beta)}}-
{{m(m-\beta)^2}\over{4(1+\beta)^2}}(K_1+1)\right] \equiv \theta^2 {{x}\over{f}}[E + F(K_1 +1)].
\end{equation}
The curvature of the surface of best focus, $\kappa_m$, is 
\begin{equation}
\label{eq:fieldcurve}
\kappa_m = {{1}\over{f}}\left[{{(m^2-2)(m-\beta) + m(m+1)}\over{m(1+\beta)}}-
{{m(m-\beta)^2}\over{2(1+\beta)^2}}(K_1+1)\right] \equiv {{1}\over{f}}[G + H(K_1 + 1)].
\end{equation}
Since the surface of best images cannot be too highly curved, the usable field of view on
the sky cannot be much larger than
\begin{equation}
\label{eq:field}
\theta_f = {{1}\over{f \kappa_m}} = [G + H(K_1 + 1)]^{-1},
\end{equation}
the field angle at which the focal surface curvature is significant.
The values of the coefficients $A$ through $H$ are listed in Table~\ref{table:constants}.

\begin{table}[hbt!]
\begin{center}
\caption[Evaluation of Aberration Coefficients for $m = -20$, $\beta = 0.1$]
{{\bf Evaluation of Aberration Coefficients for $m = -20$, $\beta = 0.1$ \ \ } 
\label{table:constants} }
\begin{tabular}{cccccccc}
\\
$A$ & $B$ & $C$ & $D$ & $E$ & $F$ & $G$ & $H$\\
%\tableline
$-0.819$ & $-14.92$ & $0.75$ & $-2877.$ & $18.19$ & $-1669.$ & $346.4$ & $3339.$\\
\end{tabular}
\end{center}
\end{table}

In the evaluation of the aberrations for visual wavelength telescopes,  it
would at this point be noted that the dominant remaining aberration is coma,
and the value of $K_1$ would be chosen so that $\theta_c = 0$ in 
Equation~\ref{eq:coma}.  This would
result in a Ritchey-Chretien design for positive $m$ and an aplanatic
Gregorian design for negative $m$.  In the aplanatic Gregorian case, $K_1$ would
be slightly larger than $-1$ ($K_1 \approx -0.95$), resulting in an ellipsoidal primary.
This eliminates third-order coma, but at the expense of a
reduced field of view, $\theta_f$.

For a submillimeter-wave telescope, however, the dominant source of blur at
the focal surface is not coma but diffraction.  The diffraction-limited
resolution on the sky is
$\theta_d = \lambda/d$, and $\theta_d > \theta_c$ for all field angles $\theta$
such that
\begin{equation}
\label{eq:diffraction}
\theta < {{\lambda f^2}\over{x^2 d}} [C + D(K_1+1)]^{-1}
\approx {{\lambda}\over{d}}
[C + D(K_1+1)]^{-1} F^2 ,
\end{equation}
where $F\equiv f/d=m f_1 / d$ is a large number.  To put it another way, coma becomes
significant compared to diffraction only for beams which are $\sim F^2/[D(K_1+1)]$ beam
diameters away from the center, and $F^2 \sim 400$.
(For visual wavelength telescopes this is a real limitation, since
the diffraction-limited beamsize is $\lambda/d \sim 2 \times 10^{-7} = 0.04''$,
and unless it is corrected, coma becomes significant $\sim 10''$ from the center of the field.)

Since diffraction swamps the classical aberrations,
the size of the field of view in the submillimeter 
will tend to be dominated by curvature of the focal surface. 
Consider, for example, the classical Gregorian case $K_1 = -1$ in 
Equation~\ref{eq:field}: $\theta_f = G^{-1} = 0.0029 = 10'$.
The size of the usable field of view can be enlarged
by setting $K_1$ so
as to maximize $\theta_f$ in Equation~\ref{eq:field}: 
\begin{equation}
\label{eq:hyperboloid}
K_1 = - {{G}\over{H}} - 1 = -1.10,
\end{equation}
resulting in a hyperboloidal primary mirror.
Substitution of this value into Equation~\ref{eq:primary} shows that the $K_1 = -1.1$ primary 
deviates from a paraboloid by $\Delta z = -50\, \rm mm$ at the outer edge of
the primary surface 
$(10300 \, \rm mm, 0, 3739 \, \rm mm)$.
Substitution into Equation~\ref{eq:conic} yields $K_2=0.67$, and
Equation~\ref{eq:secondary} shows that the secondary is then an oblate
spheroid.
Substitution in Equation~\ref{eq:astigmatism} shows that this choice worsens the 
astigmatism, but not to a degree which is significant compared to diffraction.
Substitution in Equation~\ref{eq:diffraction}, however, shows that this 
choice of primary results in
coma which is becoming important compared to diffraction.  In the fully corrected
system, where $i_1$ is chosen so that $i_0 = 0$ in Equation~\ref{eq:dragone}, the
coma will be reduced by a factor of $\sim 4$, so Equation~\ref{eq:diffraction}
should be considered an upper bound to the significance of coma.  
There is, therefore, an optimal value for
$K_1$ such that $-1.1 < K_1 < -1.0$ and the image is maximally flat and minimally
aberrated.

An intriguing possibility sets $K_2 =0$.  Then
the secondary is a concave sphere, and the exit pupil of the telescope,
which in this design is the position of the tertiary chopper, lies
near the center of that sphere.  This choice provides some
flattening of the focal surface (since $K_1 = -A/B -1 = -1.055 < -1$),
coma which is reduced compared to the $K_1 = -1.1$ case, 
and also a high degree of symmetry
at the secondary mirror during the beam chop.  This possibility
will be investigated using numerical methods.

%Following the secondary mirror are a flat tertiary (the chopper) which directs 
%the beam along the elevation axis, and a flat quaternary which 
%directs the beam down to the receiver cabin.  

\section{Construction and Management Plans} 
\begin{figure}[tbh!] \begin{center} \leavevmode
\epsfxsize=6.0in
\epsfbox{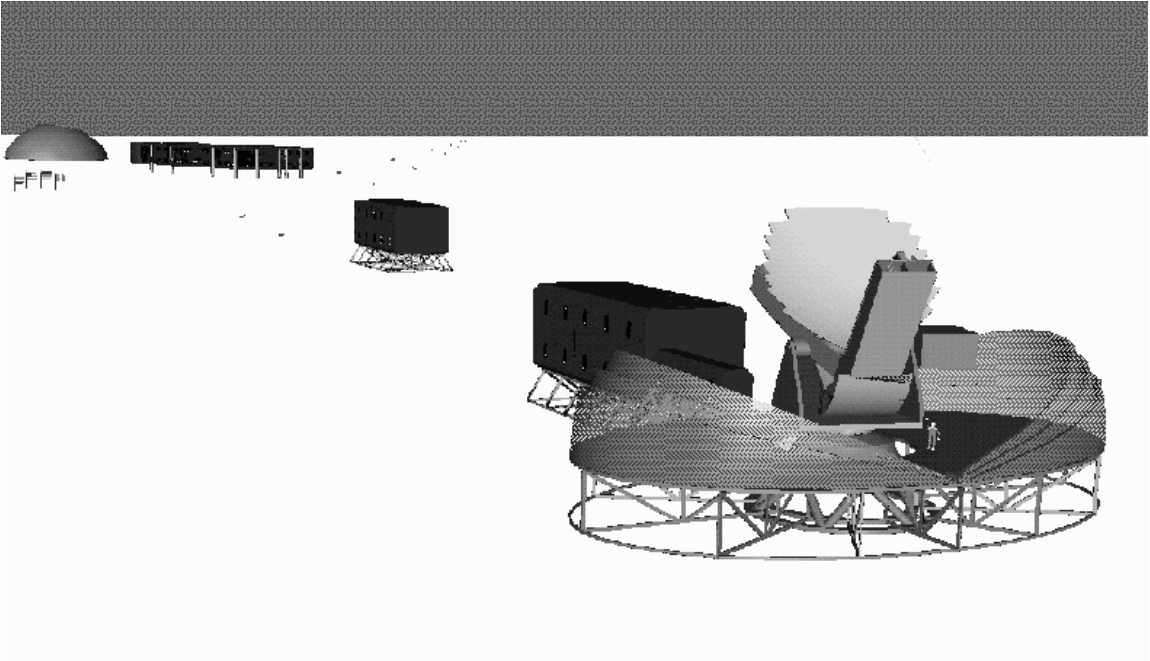}  \\
\end{center}
\caption[Computer model of the 10 meter telescope at South Pole.]
{{\bf Computer model of the 10 meter telescope at South Pole.\ \ }
The new South Pole station, the existing dome and M. A. Pomerantz Laboratory
are in the background.  All surface buildings
are elevated to reduce snow drifts.  The 10 meter support facility
will occupy one floor of the new Dark Sector laboratory building and
connect to it by a covered walkway.  A clamshell-shaped protective 
enclosure is shown in the stowed position. 
\label{fig:10matpole} }
\end{figure}

The telescope will be constructed at the Smithsonian Submillimeter Array
construction and test facility at Haystack Observatory during calendar year
2002.  From 1999 through 2001, project personnel will fully specify the 
instrument, carry out trade studies and design and performance analysis.
In the year 2003, the telescope will be tested using 
SMA test facilities.
The telescope
will be disassembled, packed, and then unpacked and
re-assembled at Haystack Observatory.  This will assure
that the third assembly at the Pole proceeds as smoothly as
possible.
The telescope and all associated equipment will be repacked
and delivered to the NSF Antarctic contractor
in Port Hueneme, California.
In 2004 and 2005, the telescope
will be reassembled and commissioned at Pole.

The 10 meter telescope will be operated as a user-facility instrument
by an international consortium.  
Observing time will be allocated with consideration for the
contributions of the international partners, but will
otherwise be freely available on a proposal basis.  The strong
survey capabilities of this instrument in the submillimeter and
for CMBR argue that a large fraction of the time (70\%) should
be allocated in large blocks to several key projects.  These
key project blocks will be allocated at three year intervals
in an open selection process involving a review of solicited
proposals.  Successful competitors
for these key project blocks would likely design and construct
detector packages.
The remaining fraction
of the observing time (30\%) will be allocated to relatively
smaller and less complex projects
which use the existing detector packages.  These projects will
be carried out by remote observing techniques, in collaboration
with SP 10m scientific staff.

%%%%%%%%%%%%%%%%%%%%%%%%%%%%%%%%%%%%%%%%%%%%%%%%%%%%%%%%%%%%%
\acknowledgments     
 
We thank P. Cheimets, D. Caldwell, W. Davis, and W. Bruckman 
for their work on the
telescope design.  We thank R. W. Wilson and A. P. Lane for their
contributions to the 10 meter proposal.
We are grateful to B. Elmegreen,
D. Fischer, P. Goldsmith, A. Lane, and G. Knapp for their
contributions to the science goals for the SP 10m.

This work was supported in part by the Smithsonian Institution and
in part by the National Science Foundation
under a cooperative agreement with the Center for Astrophysical Research
in Antarctica (CARA), grant number NSF DPP 89-20223.  CARA is a National
Science Foundation Science and Technology Center.

%%%%%%%%%%%%%%%%%%%%%%%%%%%%%%%%%%%%%%%%%%%%%%%%%%%%%%%%%%%%%
%%%%% References %%%%%

  \end{document}